\documentclass[aps,preprint,a4paper]{revtex4}%
\usepackage{amsfonts}
\usepackage{amsmath}
\usepackage{amssymb}
\usepackage{graphicx}%
\providecommand{\U}[1]{\protect \rule{.1in}{.1in}}

\begin{document}
\preprint{ }
\title[BKT in imbalanced Fermi gases]{The effect of population imbalance on the Berezinskii-Kosterlitz-Thouless
phase transition in a superfluid Fermi gas}
\author{J. Tempere$^{1,2}$}
\author{S. N. Klimin$^{1}$}
\altaffiliation{On leave of absence from: Department of Theoretical Physics, State University
of Moldova, str. A. Mateevici 60, MD-2009 Kishinev, Republic of Moldova.}

\author{J. T. Devreese$^{1}$}
\altaffiliation{Also at Technische Universiteit Eindhoven, P. B. 513, 5600 MB Eindhoven, The Netherlands.}

\affiliation{$^{1}$TFVS, Universiteit Antwerpen, B-2020 Antwerpen, Belgium.}
\affiliation{$^{2}$Lyman Laboratory of Physics, Harvard University, Cambridge, MA 02138, USA.}

\begin{abstract}
The Berezinskii-Kosterlitz-Thouless (BKT) mechanism describes the breakdown of
superfluidity in a two-dimensional Bose gas or a two-dimensional gas of paired
fermions. In the latter case, a population imbalance between the two pairing
partners in the Fermi mixture is known to influence pairing characteristics.
Here, we investigate the effects of imbalance on the two-dimensional BKT
superfluid transition, and show that superfluidity is even more sensitive to
imbalance than for three dimensional systems. Finite-temperature phase
diagrams are derived using the functional integral formalism in combination
with a hydrodynamic action functional for the phase fluctuations. This allow
to identify a phase separation region and tricritical points due to imbalance.
In contrast to superfluidity in the three-dimensional case, the effect of
imbalance is also pronounced in the strong-coupling regime.

\end{abstract}
\date{\today}
\startpage{1}
\maketitle

\section{Introduction \label{sec:intro}}

Quantum phenomena which occur at very low temperatures are a subject of
intense experimental and theoretical study. Recent progress in the
experimental investigation of ultra-cold atoms stimulated an unprecedented
interest to the theoretical problems of condensation of cold bosons and
pairing of interacting fermions (see, e. g., the review \cite{Bloch2008} and
references therein). These phenomena are related to a variety of objects
including stars, dense nuclear and quark matter, and plasma systems
\cite{Casalbuoni2004,quark,nuclear}.

Phase transitions of quantum systems strongly depend on their dimensionality.
Two-dimensional Fermi gases have remarkable features, which are not observed
in three dimensions. The quasi-2D regime for cold atoms can be reached using a
sufficiently strong confinement of atoms along one direction, so that they
occupy only the lowest size quantization subband. Advances in pairing of
trapped cold atomic Fermi gases with a controlled geometry of a trapping
potential allow experimentalists to realize systems of different
(quasi)dimensionalities. Both condensation of bosonic atoms \cite{Greiner2002}
and pairing of fermions \cite{Chin2006} has been observed recently in optical lattices.

The Mermin-Wagner-Hohenberg theorem \cite{MW} shows that in a uniform,
two-dimensional (2D) system, the long range order is destroyed by thermal
fluctuations so that 2D Bose gases cannot undergo Bose-Einstein condensation
at nonzero temperatures \cite{Bogolubov,Lifshitz,Popov}. Nevertheless, these
two-dimensional systems can form a "quasicondensate" and exhibit
superfluidity\cite{Berezinskii}. Kosterlitz and\ Thouless\cite{KT} showed that
the mechanism driving the superfluid-to-normal phase transition in this case
is the proliferation of vortices and antivortices above a critical temperature
$T_{BKT}$, spoiling the phase coherence. The experimental situation differs in
that considered by Mermin, Wagner and Hohenberg in that the gas is not
uniform. The presence of a trapping potential in the experimental realizations
changes the density of states, and in a confined 2D system, true Bose-Einstein
condensation does become possible \cite{Petrov2001}. However, the
Berezinskii-Kosterlitz-Thouless (BKT) mechanism continues to play a role in
the suppression of superfluidity also in trapped
systems\cite{theorybeforeexpt}. Indeed, a BKT-type crossover in a trapped
quantum degenerate gas of $^{87}$Rb has recently been observed by Hadzibabic
et al. \cite{Hadzibabic2006} using an interference technique to detect phase
defects\cite{phasedefects}. This has stimulated renewed theoretical interest
in the BKT transition in trapped dilute gases\cite{BKTtheorypapers}.

The discussion in the above paragraph focuses on Bose gases. Rice
\cite{Rice1965}, and Schmitt-Rink, Varma, and Ruckenstein
\cite{SshmittRink1989} extended these results to the case of a gas of paired
fermions. The pairing characteristics depend on the the interatomic
interaction strength. The superfluid phase transition in a 2D Fermi gas at
$T=0$ has been studied by Randeria \emph{et al}. \cite{Randeria1990} in a wide
range of interaction strengths from the weak-coupling limit, where the
Bardeen-Cooper-Schriffer (BCS) pairing regime is realized, to the
strong-coupling limit which corresponds to the Bose condensation of
interacting bound pairs. Petrov et al. \cite{Petrov2003} analytically
investigated pairing of a quasi 2D Fermi gas at a finite temperature in the
weak-coupling BCS limit and in the strong-coupling limit where the BKT
superfluid transition occurs. 

The influence of spin imbalance on the superfluid properties of a Fermi gas is
of particular interest. A phase separation between the superfluid and normal
component of an interacting Fermi gas of cold atoms with unequal spin
populations in 3D was observed in Refs.
\cite{Zwierlein2006,Partridge2006,Shin2008}. In these experiments, both the
interaction strength between ultra-cold fermions and the population imbalance
can be controlled using the Feshbach resonance. Allowing the control of
population or mass imbalance opens a unique possibility to investigate the
stability of fermion pairing and, in particular, to determine experimentally
the equation of state for an imbalanced fermion system \cite{Shin2008}.

The BKT phase transition for a balanced 2D Fermi gas with the $s$-wave
scattering in the balanced case has been theoretically analyzed by Botelho and
de Melo \cite{Botelho2006}. They treated the fermion pairing in 2D by the path
integral technique taking into account phase fluctuations. Within a similar
approach, using an effective Hamiltonian which involves fermions interacting
with each other and with dressed molecules, the BKT transition has been
considered for a quasi-2D trapped Fermi gas \cite{Zhang2008}.

In the present work, we extend the approach of Ref. \cite{Botelho2006} to
investigate the effect of population imbalance on the BKT phase transition in
a 2D Fermi gas. Using the Hubbard-Stratonovich transformation, we derive a
hydrodynamic effective bosonic action, which in the limiting case of a
balanced gas is reduced to the effective action of Ref. \cite{Botelho2006}. On
the basis of the obtained effective action, we analyze phase diagrams for an
interacting, imbalanced Fermi gas in 2D. The paper is organized as follows. In
Sec. \ref{sec:theory}, we describe the theoretical formalism for interacting
fermions in 2D. In Sec. \ref{sec:results}, we analyze the dependence of the
critical temperature of the BKT transition on the coupling strength and on the
population imbalance. The section is followed by conclusions, Sec.
\ref{sec:conclusions}.

\section{Functional integral description\label{sec:theory}}

\subsection{General formalism}

We consider a two-component gas of interacting fermions in 2D, with the
$s$-wave pairing and with a population imbalance. The partition function of
the system of fermions in 2D is expressed as the path integral over Grassmann
variables $\left[  \bar{\psi}_{\mathbf{x},\tau,\sigma},\psi_{\mathbf{x}%
,\tau,\sigma}\right]  $,%
\begin{equation}
\mathcal{Z}\propto \int \mathcal{D}\left[  \bar{\psi}_{\mathbf{x},\sigma}\left(
\tau \right)  ,\psi_{\mathbf{x},\sigma}\left(  \tau \right)  \right]
\exp \left(  -S\right)  .
\end{equation}
The action functional%
\begin{equation}
S=S_{0}+S_{int} \label{S}%
\end{equation}
is a sum of the free-fermion and interaction contributions,%
\begin{align}
S_{0}  &  =\int_{0}^{\beta}d\tau \int d^{2}\mathbf{x}\sum_{\sigma
=\uparrow,\downarrow}\left[  \bar{\psi}_{\mathbf{x},\sigma}\left(
\tau \right)  \left(  \frac{\partial}{\partial \tau}-\nabla_{\mathbf{x}}^{2}%
-\mu_{\sigma}\right)  \psi_{\mathbf{x},\sigma}\left(  \tau \right)  \right]
,\label{ff}\\
S_{int}  &  =\int_{0}^{\beta}d\tau \int d^{2}\mathbf{x}\int d^{2}%
\mathbf{y}V\left(  \mathbf{x}-\mathbf{y}\right)  \bar{\psi}_{\mathbf{x}%
,\uparrow}\left(  \tau \right)  \bar{\psi}_{\mathbf{y},\downarrow}\left(
\tau \right)  \psi_{\mathbf{y},\downarrow}\left(  \tau \right)  \psi
_{\mathbf{x},\uparrow}\left(  \tau \right)  , \label{int}%
\end{align}
where $\beta=\frac{1}{k_{B}T}$ is the inverse to the temperature. We use the
units in which $\hbar=1$, the fermion mass $m=1/2$, and the Fermi energy
$E_{F}\equiv \left(  2\pi n_{0}\right)  ^{2/3}/\left(  2m\right)  =1$ (where
$n_{0}$ is the fermion density in 2D). We express the results below in terms
of the averaged chemical potential $\mu=\left(  \mu_{\uparrow}+\mu
_{\downarrow}\right)  /2$ determining the total number of fermions and the
imbalance potential $\zeta=\left(  \mu_{\uparrow}-\mu_{\downarrow}\right)
/2$. For the interaction potential, we use the separable expression proposed
in Refs. \cite{Botelho2006,Duncan2002},%
\begin{equation}
V_{\mathbf{k,k}^{\prime}}=g\Gamma_{\mathbf{k}}\Gamma_{\mathbf{k}^{\prime}},
\label{sep}%
\end{equation}
where $g$ is the interaction strength. The factor $\Gamma_{\mathbf{k}}$
describes a finite-range potential,%
\begin{equation}
\Gamma_{\mathbf{k}}=\left(  1+\frac{k}{k_{0}}\right)  ^{-1/2}, \label{g}%
\end{equation}
where $R\sim k_{0}^{-1}$ plays the role of the interaction range. The
particular case of the contact interaction corresponds to $k_{0}%
\rightarrow \infty$ so that $\Gamma_{\mathbf{k}}\rightarrow1$. The interaction
term $S_{int}$ of the action functional is then given by%
\begin{equation}
S_{int}=\int_{0}^{\beta}d\tau \frac{g}{L^{2}}\sum_{\mathbf{q}}\bar
{B}_{\mathbf{q}}\left(  \tau \right)  B_{\mathbf{q}}\left(  \tau \right)  .
\end{equation}
where $L$ is the lateral size of the 2D system, and the collective coordinates
$B_{\mathbf{q}}\left(  \tau \right)  $ are determined as%
\[
B_{\mathbf{q}}\left(  \tau \right)  \equiv \sum_{\mathbf{k}}\Gamma_{\mathbf{k}%
}a_{-\mathbf{k}+\frac{\mathbf{q}}{2},\downarrow}\left(  \tau \right)
a_{\mathbf{k}+\frac{\mathbf{q}}{2},\uparrow}\left(  \tau \right)  .
\]
Further on, we apply the Hubbard-Stratonovich (HS) transformation. Introducing
the extended action%
\begin{equation}
S_{ext}=S-\int_{0}^{\beta}d\tau \frac{1}{g}\sum_{\mathbf{q}}\bar{\phi
}_{\mathbf{q}}\left(  \tau \right)  \phi_{\mathbf{q}}\left(  \tau \right)
\label{Sext}%
\end{equation}
with the auxiliary Bose field (HS field) $\phi_{\mathbf{q}}\left(
\tau \right)  $, and performing the shift of boson coordinates, which
eliminates the fermion-fermion interaction term $S_{int}$, the HS
transformation results in the action%
\begin{align}
S_{ext}  &  =S_{0}-\int_{0}^{\beta}d\tau \frac{1}{L}\sum_{\mathbf{q}}\left[
\bar{B}_{\mathbf{q}}\left(  \tau \right)  \phi_{\mathbf{q}}\left(  \tau \right)
+B_{\mathbf{q}}\left(  \tau \right)  \bar{\phi}_{\mathbf{q}}\left(
\tau \right)  \right] \nonumber \\
&  -\int_{0}^{\beta}d\tau \frac{1}{g}\sum_{\mathbf{q}}\bar{\phi}_{\mathbf{q}%
}\left(  \tau \right)  \phi_{\mathbf{q}}\left(  \tau \right)  . \label{Sext0}%
\end{align}

Because the phase fluctuations about the saddle point are, in general, not
small, the boson (HS) and fermion variables in the coordinate representation
are transformed as \cite{DePalo1999}:%
\begin{equation}
\phi_{\mathbf{r}}\left(  \tau \right)  =\phi_{\mathbf{r}}^{\prime}\left(
\tau \right)  e^{i\theta_{\mathbf{r}}\left(  \tau \right)  },\; \psi
_{\mathbf{r},\sigma}\left(  \tau \right)  =\psi_{\mathbf{r},\sigma}^{\prime
}\left(  \tau \right)  e^{i\theta_{\mathbf{r}}\left(  \tau \right)  /2}.
\label{Trans}%
\end{equation}

In the same formalism for a Fermi gas in 3D, the further step is the path
integration over fermion variables and the expansion of the resulting bosonic
action over fluctuations about the saddle point \cite{deMelo1993}. This method
provides a description of a superfluid phase transition between the normal
phase and the true condensate of fermion pairs. In the 2D case, at least when
restricting the expansion by quadratic fluctuations, the superfluidity occurs
only at $T=0$ \cite{SshmittRink1989}. Traven \cite{Traven1994} showed that the
interaction between fluctuations of the pairing field in a 2D attractive Fermi
gas allows a superfluid phase transition at a very low temperature. However,
the superfluid state can exist in a 2D Fermi gas at relatively high
temperatures as a quasicondensate, i. e., a state with fermion pairing
\textquotedblleft where the density fluctuations are suppressed but the phase
still fluctuates\textquotedblright \  \cite{Petrov2000}. In two dimensions, the
quasicondensate can be realized through bound vortex-antivortex pairs
\cite{Bloch2008}.

\subsection{Fluctuating phase}

After integrating out the fermion (Grassmann) variables, an affective action
in the bose field $\phi_{\mathbf{r}}$ is obtained. The remaining functional
integral over this Bose field cannot be taken in general. Several levels of
approximation can be made to get results. The crudest approximation is the
mean-field approximation which replaces the field by a constant,
$\phi_{\mathbf{r}}\rightarrow \left \vert \Delta \right \vert $, the saddle point.
To improve on this, fluctuations around the saddle point can still be taken
into account; this can be done in an exact way only up to quadratic order in
the fluctuation. One can choose to write the fluctuations as $\phi
_{\mathbf{r}}\rightarrow \left \vert \Delta \right \vert +\delta_{\mathbf{r}}$,
with $\delta_{\mathbf{r}}$ complex; this corresponds to the NSR approach that
investigates the presence of a real condensate rather than a quasicondensate.
Alternatively the NSR fluctuations can be written as amplitude and phase
fluctuations, $\phi_{\mathbf{r}}\rightarrow \left(  \left \vert \Delta
\right \vert +\left \vert \delta_{\mathbf{r}}\right \vert \right)  e^{i\theta
_{\mathbf{r}}}$ where $\left \vert \delta_{\mathbf{r}}\right \vert $ and
$e^{i\theta_{\mathbf{r}}}$ are real fields. In order to treat the
quasicondensate state, relevant for the 2D case, we have to focus on phase
fluctuations: that corresponds to setting $\phi_{\mathbf{r}}\rightarrow
\left \vert \Delta \right \vert e^{i\theta_{\mathbf{r}}}$. We will simplify the
notation and write $\left \vert \Delta \right \vert =\Delta$ for the energy gap
parameter of the Bogoliubov excitations. Moreover, we will assume that the
remaining fluctuation field $\theta_{\mathbf{r}}$ varies slowly as a function
of position and time with respect to the variations of the fermion fields. A
similar assumption was used in Refs. \cite{DePalo1999,Zhang2008}.\ The
resulting hydrodynamic action is structurally similar to the saddle-point
action for imbalanced fermions in 2D \cite{Tempere2007}
\begin{equation}
S_{eff}=-\int_{0}^{\beta}d\tau \int d^{2}\mathbf{r}\frac{1}{L^{2}}%
\sum_{\mathbf{k}}\left[  \frac{1}{\beta}\ln \left(  2\cosh \beta \zeta
_{\mathbf{k}}+2\cosh \beta \tilde{E}_{\mathbf{k}}\right)  -\tilde{\xi
}_{\mathbf{k}}\right]  -\frac{\beta L^{2}}{g}\Delta^{2}, \label{Sef}%
\end{equation}
in which the fermion energy $\xi_{\mathbf{k}}=k^{2}-\mu$, the Bogolubov
excitation energy $E_{\mathbf{k}}=\sqrt{\xi_{\mathbf{k}}^{2}+\Delta^{2}%
\Gamma_{\mathbf{k}}^{2}}$ and the imbalance potential $\zeta$ are replaced by
expressions depending on boson coordinates:%
\begin{align}
\tilde{\xi}_{\mathbf{k}}  &  =k^{2}-\tilde{\mu},\; \tilde{E}_{\mathbf{k}%
}=\sqrt{\tilde{\xi}_{\mathbf{k}}^{2}+\Delta^{2}\Gamma_{\mathbf{k}}^{2}%
},\label{p1}\\
\tilde{\mu}  &  =\mu-\frac{i}{2}\frac{\partial \theta}{\partial \tau}-\frac
{1}{4}\left(  \nabla \theta \right)  ^{2}+\frac{i}{2}\nabla^{2}\theta
,\label{p2}\\
\zeta_{\mathbf{k}}  &  =\zeta-\nabla \theta \cdot \mathbf{k}. \label{p3}%
\end{align}

Keeping the quadratic-order fluctuation terms we arrive at the action
$S_{eff}$ as the sum of the saddle-point action $S_{sp}$, which coincides with
the expression (5) of Ref. \cite{Tempere2007}, and the diagonal quadratic form
of the phase fluctuations%
\begin{equation}
S_{fl}=\frac{1}{2}\int_{0}^{\beta}d\tau \int d^{2}\mathbf{r}\left[  A\left(
\frac{\partial \theta}{\partial \tau}\right)  ^{2}+\rho_{s}\left(  \nabla
\theta \right)  ^{2}\right]  . \label{sfl5}%
\end{equation}
The coefficients in the fluctuation action are the pair superfluid density%
\begin{equation}
\rho_{s}=\frac{1}{L^{2}}\sum_{\mathbf{k}}\left[  \frac{1}{2}\left(
1-\frac{\xi_{k}}{E_{k}}X_{k}\right)  -\frac{k^{2}}{2}Y_{\mathbf{k}}\right]
\label{rs}%
\end{equation}
and the constant%
\[
A=\frac{1}{4L^{2}}\sum_{\mathbf{k}}\left(  \frac{\Gamma_{k}^{2}\Delta^{2}%
}{E_{\mathbf{k}}^{3}}X_{k}+\frac{\xi_{k}^{2}}{E_{k}^{2}}Y_{k}\right)  ,
\]
where $X_{k}$ the function%
\[
X_{k}=\frac{\sinh \beta E_{k}}{\cosh \beta \zeta+\cosh \beta E_{k}}%
\]
and $Y_{k}$ is the extension of the Yoshida distribution to imbalanced
fermions:%
\begin{equation}
Y_{k}=\beta \frac{\cosh \beta \zeta \cosh \beta E_{k}+1}{\left(  \cosh \beta
\zeta+\cosh \beta E_{k}\right)  ^{2}}.
\end{equation}

The action in Eq. (\ref{sfl5}) describes a Bose gas of spin
waves\cite{Botelho2006} with the energy spectrum $\omega_{k}$ given by
\begin{equation}
\omega_{k}=ck,\;c=\sqrt{\frac{\rho_{s}}{A}}.
\end{equation}
The spin wave contribution to the thermodynamic potential is given by the
expression%
\begin{equation}
\Omega_{sw}=\frac{1}{\beta}\sum_{\mathbf{k}}\ln \left(  1-e^{-\beta \omega_{k}%
}\right)  . \label{W}%
\end{equation}
The total thermodynamic potential of the fermion system taking into account
phase fluctuations is the sum of the spin-wave term (\ref{W}) and the
saddle-point contribution, which is provided by the saddle-point action,%
\begin{equation}
\Omega_{sp}=\sum_{\mathbf{k}}\left[  \frac{1}{\beta}\ln \left(  2\cosh
\beta \zeta+2\cosh \beta E_{\mathbf{k}}\right)  -\xi_{\mathbf{k}}\right]
-\frac{\beta L^{2}}{g}\Delta^{2}. \label{Wsp}%
\end{equation}
For an imbalanced quasi 2D Fermi gas in an optical potential, the mean-field
zero-temperature phase diagrams were analyzed in Ref. \cite{Tempere2007} on
the basis of this saddle-point action (neglecting spin-wave contributions).

The gap parameter $\Delta$ for an imbalanced Fermi gas is determined through
the minimization of the saddle-point thermodynamic potential $\Omega_{sp}$ as
a function of $\Delta$ at given $\beta,\mu,\zeta$:%
\begin{equation}
\left(  \frac{\partial \Omega_{sp}}{\partial \Delta}\right)  _{\beta,\mu,\Delta
}=0. \label{min}%
\end{equation}
For a complete determination of thermodynamic parameters at a given
temperature, the minimum condition (\ref{min}) is solved jointly with the
number equations:%
\begin{equation}
n\equiv-\left(  \frac{\partial}{\partial \mu}\frac{\Omega}{L^{2}}\right)
_{\beta,\zeta,\Delta}=\frac{1}{2\pi},\; \delta n\equiv-\left(  \frac{\partial
}{\partial \zeta}\frac{\Omega}{L^{2}}\right)  _{\beta,\mu,\Delta}=\frac{1}%
{2\pi}\frac{\delta n}{n}, \label{number}%
\end{equation}
where $n$ and $\delta n$ are the total fermion density $n$ and the density
difference $\delta n$, respectively.

\subsection{BKT transition temperature}

The coupled gap and number equations (\ref{min}) and (\ref{number}) have to be
solved together. In equation (\ref{number}), three different levels of
approximation can be made, in analogy to the approximations for the bose field
$\phi_{\mathbf{r}}(\tau)$ as discussed in the beginning of the previous subsection.

The first and simplest case is the mean-field approximation (as described in
Refs. \cite{Botelho2006,Zhang2008}), where we use $\Omega=\Omega_{sp}$. This
corresponds to setting the (Hubbard-Stratonovic) Bose field equal to a
constant $\phi=\Delta_{MF}$ (both constant in amplitude and in phase). The
constant can be determined by minimizing $\Omega_{sp}$ and allows to determine
the phase transition line between the normal state, in which $\Delta_{MF}=0$,
and the quasicondensate of fermion pairs, in which $\Delta_{MF}\neq0$. The
temperature separating the aforementioned phases will be denoted by $T_{MF}$.
For a balanced gas, the phase transition between normal and paired states is
of the second order. For an imbalanced gas, also the first-order phase
transition between the normal and paired states is possible.

The paired state below $T_{MF}$ is not always superfluid. Proliferating
vortices and antivortices destroy phase coherence in the quasicondensate, and
suppress superfluidity\cite{KT}. The mean-field approximation does not allow
to determine the temperature $T_{BKT}$ separating the superfluid
quasicondensate state from the non-superfluid paired state. To investigate the
question of superfluidity and determine the BKT temperature, we need to use
$\Omega=\Omega_{sp}+\Omega_{sw}$ in (\ref{number}), where $\Omega_{sw}$ is the
free energy contribution (\ref{W}) from the phase fluctuations. This
corresponds to giving the (Hubbard-Stratonovic) Bose field a constant
amplitude, but allowing its phase to fluctuate freely. The BKT transition
temperature is then a root of the universal Nelson-Kosterlitz equation
\cite{Nelson1977}%
\begin{equation}
T_{BKT}-\frac{\pi}{2}\rho_{s}\left(  T_{BKT}\right)  =0. \label{TKT}%
\end{equation}
As distinct from the aforesaid phase transition at $T=T_{MF}$, the BKT phase
transition at $T=T_{BKT}$ is characterized by an abrupt change of the
superfluid density from zero to a finite value satisfying Eq. (\ref{rs}). The
phase-fluctuation contribution to the density vanishes at the phase boundary
between the paired state and the normal state, because at $\Delta_{MF}=0$ the
superfluid density $\rho_{s}$ turns to zero.

Finally, the approximation of setting $\Omega=\Omega_{sp}+\Omega_{sw}$ in
(\ref{number}) does not allow to determine whether the superfluid
quasicondensate can form a true condensate below some critical temperature
$T_{BEC}$. Note that the emergence of a true condensate is only possible in a
finite system, as $T_{BEC}\rightarrow0$ for $L\rightarrow \infty$. To determine
the critical temperature we should use $\Omega=\Omega_{sp}+\Omega_{flucts}$ in
(\ref{number}), where $\Omega_{flucts}$ contains contributions from \emph{both
}amplitude and phase fluctuations in the Hubbard-Stratonovic Bose field.

\section{Results and discussion\label{sec:results}}

When $T_{MF}>0$, equation (\ref{TKT}) can be satisfied if the superfluid
density $\rho_{s}\neq0$. At $T=T_{MF}$, the superfluid density for a balanced
Fermi gas turns to zero. Therefore for a balanced Fermi gas $T_{BKT}$ is
always lower than $T_{MF}$. On the contrary, in the case of unequal
\textquotedblleft spin up\textquotedblright \ and \textquotedblleft spin
down\textquotedblright \ populations, for a sufficiently high population
imbalance the phase transition at $T=T_{MF}$ can be of the first order, when
the gap parameter $\Delta$ changes discontinuously from zero to a finite
value. In this case, as shown below, $T_{BKT}$ and $T_{MF}$ can coincide in a
definite range of the coupling strength.

In the region between $T_{BKT}$ and $T_{MF}$, the superfluidity is destroyed
owing to a proliferation of free vortices. However, the phase of a 2D Fermi
gas in this region is not a uniform normal state, because the gap parameter
for $T_{BKT}<T<T_{MF}$ can be other than zero, so that fermion pairing can
occur in that temperature region. The region between $T_{BKT}$ and $T_{c}$ can
be therefore attributed to a state in which pairing can occur but the phase
coherence is destroyed. For slightly higher temperatures with respect to
$T_{BKT}$, vortices form a disordered gas of phase defects \cite{Bloch2008}.
For higher temperatures the concept of vortices is inapplicable due to density fluctuations.

In general, there are the following regions in phase diagrams for an
imbalanced Fermi gas in 2D: (1) the normal state in which the gap parameter
$\Delta=0$, (2) the superfluid quasicondensate state in which $\Delta \neq0$
and $\rho_{s}\neq0$, (3) the state in which pairing is possible but without
phase coherence, and (4) the phase separation region, which appears due to a
population imbalance. In the latter one, no solution exist for the set of
equations (\ref{min}) and (\ref{number}). Therefore, in the phase separation
region the system will unmix in a phase with lower\ (or no) imbalance and the
normal phase at higher imbalance. At the first-order phase transition, the
system abruptly passes through the phase separation region.

\bigskip

In Fig. 1, we represent phase diagrams in the variables $\left \{
T,E_{b}\right \}  $, where $T$ is the temperature, and $E_{b}$ is the binding
energy of a two-particle bound state in 2D. The phase diagrams in Fig. 1 are
obtained at a given value of the imbalance chemical potential $\zeta$. The top
panel shows the balanced case ($\zeta=0$), and the middle and lower panels
illustrate how the diagram changes with increasing difference between the
chemical potentials of up and down species. The energy $E_{b}$ is the
parameter which characterizes the coupling strength of the fermion-fermion
attractive interaction. In two dimensions, the strength $g$ of the contact
interaction is related to this binding energy through\cite{Randeria1990}.%
\begin{equation}
\frac{1}{g}=\frac{m}{4\hbar^{2}}\left[  i-\frac{\ln \left(  E/E_{b}\right)
}{\pi}\right]  .
\end{equation}
The binding energy itself can be related to the experimental parameters. The
two-dimensional system is created through a strong confinement of the third
direction; in general this strong confinement can be associated with a
harmonic potential with frequency $\omega_{L}$ (and oscillator length
$\ell_{L}).$ The two-particle bound state exists for all values of the (3D)
s-wave scattering length $a_{s}$ of the fermionic atoms and its binding energy
is given by:%
\begin{equation}
E_{b}=\frac{C\hbar \omega_{L}}{\pi}\exp \left(  \sqrt{2\pi}\frac{\ell_{L}}%
{a_{s}}\right)  ,
\end{equation}
with $C\approx0.915$ (cf. Ref. \cite{PetrovPRA64}).

The dashed curves in Fig. 1 correspond to the phase transition curves,
$T_{MF}(E_{b})$, within the mean-field approach. When the system is
imbalanced, two changes occur with respect to the balanced case: (i) below a
certain value of the binding energy, $T_{MF}\rightarrow0$, and (ii) a
tricritical point appears on this curve. The mean-field phase transition curve
splits below this tricritical point and the phase separation region opens up.
In this tricritical point, there is coexistence of the paired state without
phase coherence, the normal state, and the phase separation region. The area
of the phase separation region is narrow and broadens with increasing the
imbalance potential.

The full curve corresponds to the BKT phase transition line $T_{BKT}(E_{b})$.
For the balanced case, we retrieve the results obtained in Ref.
\cite{Botelho2006} and find that the BKT transition temperature is nonzero for
any finite binding energy. In particular, we find that in the strong-coupling
limit (for large $E_{b}$), the superfluid density $\rho_{s}$ rises to its
maximum, equal to half the total fermion density, so that the BKT transition
temperature obtained from Eq. (\ref{TKT}) tends to a finite value
$T_{BKT}^{\left(  \max \right)  }=(1/8)T_{F}$. When imbalance is introduced,
two qualitative changes occur in the BKT transition curve. Firstly, below a
critical value of the binding energy, $T_{BKT}\rightarrow0$. Imbalance
suppresses BKT-superfluidity as it does for 3D-superfluidity. The critical
binding energy needed to restore superfluidity shifts to higher values as the
imbalance potential grows. Secondly, there appears a tricritical point also on
the BKT curve, indicating coexistence of the superfluid state, the state with
pairing but no phase coherence, and the phase separation region.

The dotted curve in the region of the normal state separates two regimes: (i)
the regime where the thermodynamic potential contains only one minimum at
$\Delta=0$ above the dotted curve, and (ii) the regime where the thermodynamic
potential contains two minima: the lower minimum at $\Delta=0$ and a higher
one at a value $\Delta \neq0$. This second minimum suggests the existence of a
metastable superfluid state.

\bigskip

In Fig. 2, we choose to fix $\delta n/n$, the relative population imbalance
itself, rather than the imbalance potential $\zeta$, and study the phase
diagram in the variables $\left \{  E_{b},T,\frac{\delta n}{n}\right \}  $. The
inset shows a slice of this phase diagram at $E_{b}=0.01$. At this binding
energy, the BKT transition temperature is a non-single-valued function of
$\delta n/n$. This result is not unexpected, because a similar
non-single-valued behavior of the critical temperature exists also for the
superfluid phase transition in a 3D Fermi gas \cite{Parish2007}. In the 3D
case this can be related to the Sarma mechanism\cite{SarmaJPCS24}: at nonzero
temperatures, a balanced superfluid system coexists with an imbalanced gas of
thermal excitations -- these excitations carry some of the excess majority
component of the imbalanced gas.

The BKT phase transition for a Fermi gas in 2D is suppressed at lower values
of the population imbalance than the superfluid phase transition for a Fermi
gas in 3D. Furthermore, this suppression is more strongly revealed at higher
binding energies. The reason of such a dependence of the BKT phase transition
as a function of the binding energy consists in the following. Let us consider
first a part of the phase diagram in the variables $\left(  E_{b}%
,T,\frac{\delta n}{n}\right)  $ taking into attention only the mean-field
phase transition. Similar phase diagrams were calculated for a 3D Fermi gas in
Ref. \cite{Parish2007}. An important feature of phase diagrams for both 3D and
2D imbalanced Fermi gases is the fact that the phase separation region lies at
lower temperatures with respect to the superfluid state. This means that, when
the temperature is gradually lowered at a fixed imbalance $\delta n/n$ (which
is assumed to be sufficiently low to pass the region of the superfluid state),
we can observe the superfluid phase transition from the normal state to the
superfluid state, and then, when the temperature is lowered further, the
superfluid state becomes non-stable in the phase separation region. So, a
sufficient lowering of the temperature leads to an instability of the
superfluid state of an imbalanced Fermi gas. On the one hand, with increasing
the coupling strength, the phase separation region broadens, and its
upper-temperature boundary shifts to higher temperatures. On the other hand,
the BKT transition temperature does not unrestrictedly increase with an
increasing coupling strength. As follows from the Nelson-Kosterlitz equation
(\ref{TKT}), the upper bound for $T_{BKT}$ is $T_{BKT}=(1/8)T_{F}$. This is
the case when the superfluid pair density achieves its maximal possible value
for a balanced gas $\rho_{s}^{\left(  \max \right)  }=\frac{1}{2}n=\frac
{1}{4\pi}$. As a result, with an increasing coupling strength, we arrive at a
stronger overlap between the area of the superfluid quasicondensate state
(below $T_{BKT}$) and the phase separation region. It is natural to expect
that at sufficiently high binding energy, the phase separation region can
completely cover the area below $T_{BKT}$, so that the superfluid state
vanishes at all temperatures. In this connection, the study of phase diagrams
for a 2D Fermi gas in the variables $\left(  T,E_{b}\right)  $ at a given
relative population imbalance (i. e., slices of the above considered
three-dimensional phase diagram at fixed $\frac{\delta n}{n}$) represents a
particular interest.

In Fig. 3, we have plotted phase diagrams for a 2D Fermi gas in the variables
$\left(  T,E_{b}\right)  $ at the relative population imbalance $\frac{\delta
n}{n}=0.03$ and 0.1. We see that even at a low population imbalance
$\frac{\delta n}{n}=0.03$, the region where the superfluid state exists is
relatively narrow. Furthermore, with an increasing binding energy $E_{b}$, the
upper-temperature bound of the phase separation region rises faster than the
BKT transition temperature. As a result, the range of coupling strengths at
which the BKT superfluid phase transition can occur, is restricted both from
lower and higher coupling. For the comparison, the regular superfluid phase
transition in a 3D imbalanced Fermi gas is restricted only from lower
couplings \cite{Tempere2008}. We can see from Fig. 3 (\emph{b}) that for a
higher (but relatively low) imbalance $\frac{\delta n}{n}=0.1$, the
higher-temperature bound of the phase separation region lies higher than
$T_{BKT}$ for all considered values of the binding energy $E_{b}$. As a
result, for $\frac{\delta n}{n}=0.1$ the superfluid state is absent. The
obtained behavior of phase diagrams for a 2D Fermi gas confirms our
suggestions above.

\section{Conclusions \label{sec:conclusions}}

In summary, we have described the effects of imbalance on the
Berezinskii-Kosterlitz-Thouless superfluid transition in a 2D Fermi gas,
through the functional integral formalism. Owing to a population imbalance,
the superfluid state cannot exist for all values of the coupling strength, but
only above a certain critical binding energy which depends on the imbalance.
The larger the imbalance potential, the higher is this critical binding
energy. As distinct from the balanced case, there is a phase-separated region
in phase diagrams, in which no uniform state can exist. As a result,
tricritical points appear at the phase diagrams, in which three different
regimes coexist. The area of the phase-separated region increases with
increasing the population imbalance. Due to the rise of the upper-temperature
bound of the instability region, the area of the superfluid state at a fixed
relative population imbalance decreases with increasing the binding energy.
Therefore a population imbalance is a factor destroying superfluidity in 2D
systems, especially at high binding energies. The BKT transition can be
experimentally observable for an imbalanced Fermi gas through the phase
separation in a quasi 2D trap. As follows from the obtained results, the
parameters of the state of that system (e. g., critical temperatures and/or
density profiles) are expected to be substantially more sensitive to the
population imbalance than the corresponding parameters for a 3D gas.

\begin{acknowledgments}
The authors would like to acknowledge fruitful discussions with M. Wouters, M.
Oberthaler, and H.T.C. Stoof. This work was supported by FWO-V projects
G.0115.06, G.0356.06, G.0180.09N, G.0370.09N, the WOG WO.033.09N (Belgium),
and INTAS Project no. 05-104-7656. J.T. gratefully acknowledges support of the
Special Research Fund of the University of Antwerp, BOF NOI UA 2004.
\end{acknowledgments}

\newpage

\begin{quote}
\textbf{Figure captions}
\end{quote}

Fig. 1. Phase diagrams for a 2D imbalanced Fermi gas in the variables $\left(
T,E_{b}\right)  $ for different values of the imbalance potential $\zeta$. The
white area is the normal state, the blue area is the superfluid state, the
green area is the state with pairing and without phase coherence, and the grey
area is the phase separation region. The dotted curve indicates the BKT
transition to a metastable superfluid state.

\bigskip

Fig. 2. Three-dimensional phase diagram for a 2D imbalanced Fermi gas in the
variables $\left(  E_{b},T,\frac{\delta n}{n}\right)  $ The slice at
$E_{b}=0.01$ shows the phase diagram in the variables $\left(  T,\frac{\delta
n}{n}\right)  $ with the same denotations as in Fig. 1.

\bigskip

Fig. 3. Phase diagram for an imbalanced 2D gas of cold fermions in the
variables $\left(  T,E_{b}\right)  $ at the relative population imbalance
$\frac{\delta n}{n}=0.03$ (\emph{a}) and $\frac{\delta n}{n}=0.1$ (\emph{b}).
The denotations are the same as in Fig. 1.

\newpage

\begin{quote}
\textbf{Figures}
\end{quote}

%

\begin{figure}
[h]
\begin{center}
\includegraphics[
height=7.2644in,
width=3.2404in
]%
{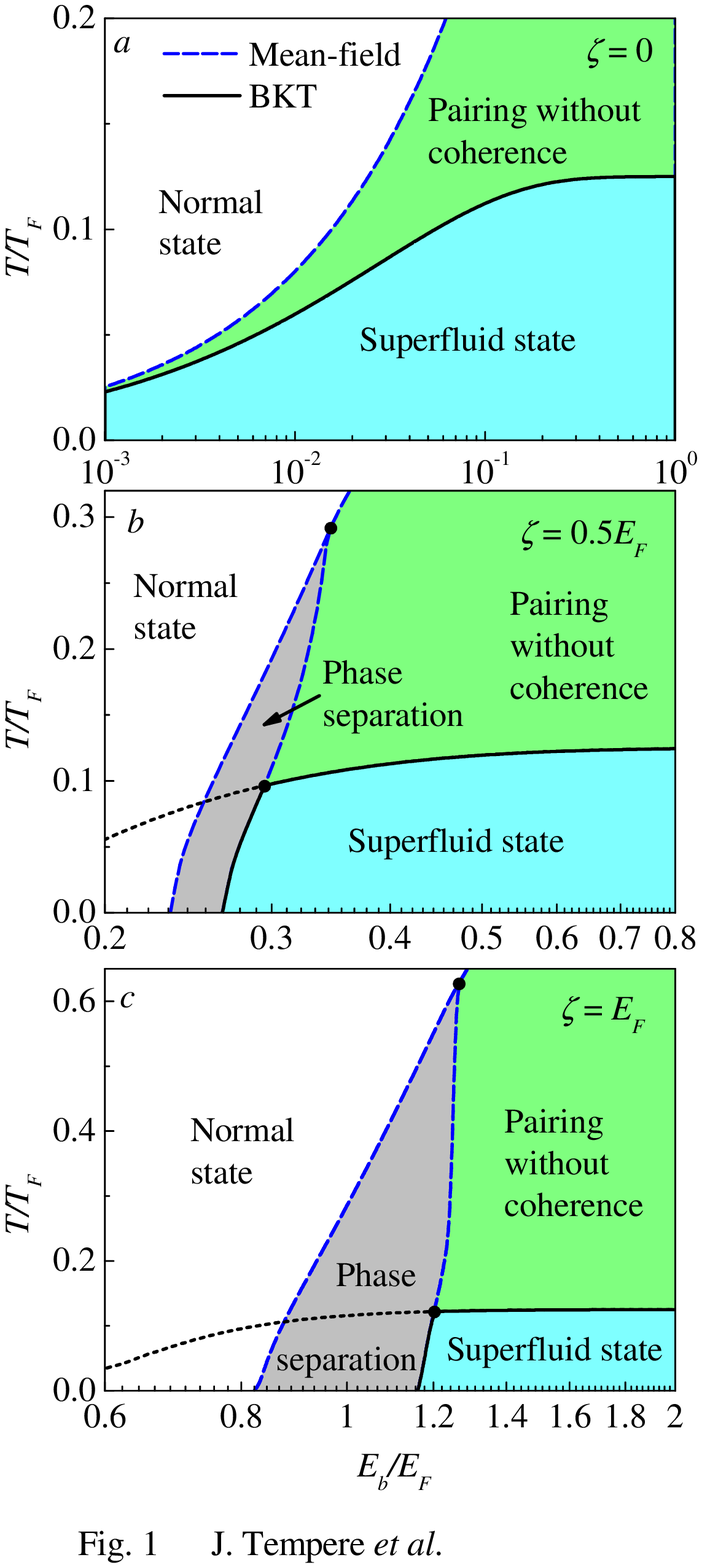}%
\end{center}
\end{figure}

\newpage%

\begin{figure}
[h]
\begin{center}
\includegraphics[
height=5.847in,
width=4.2973in
]%
{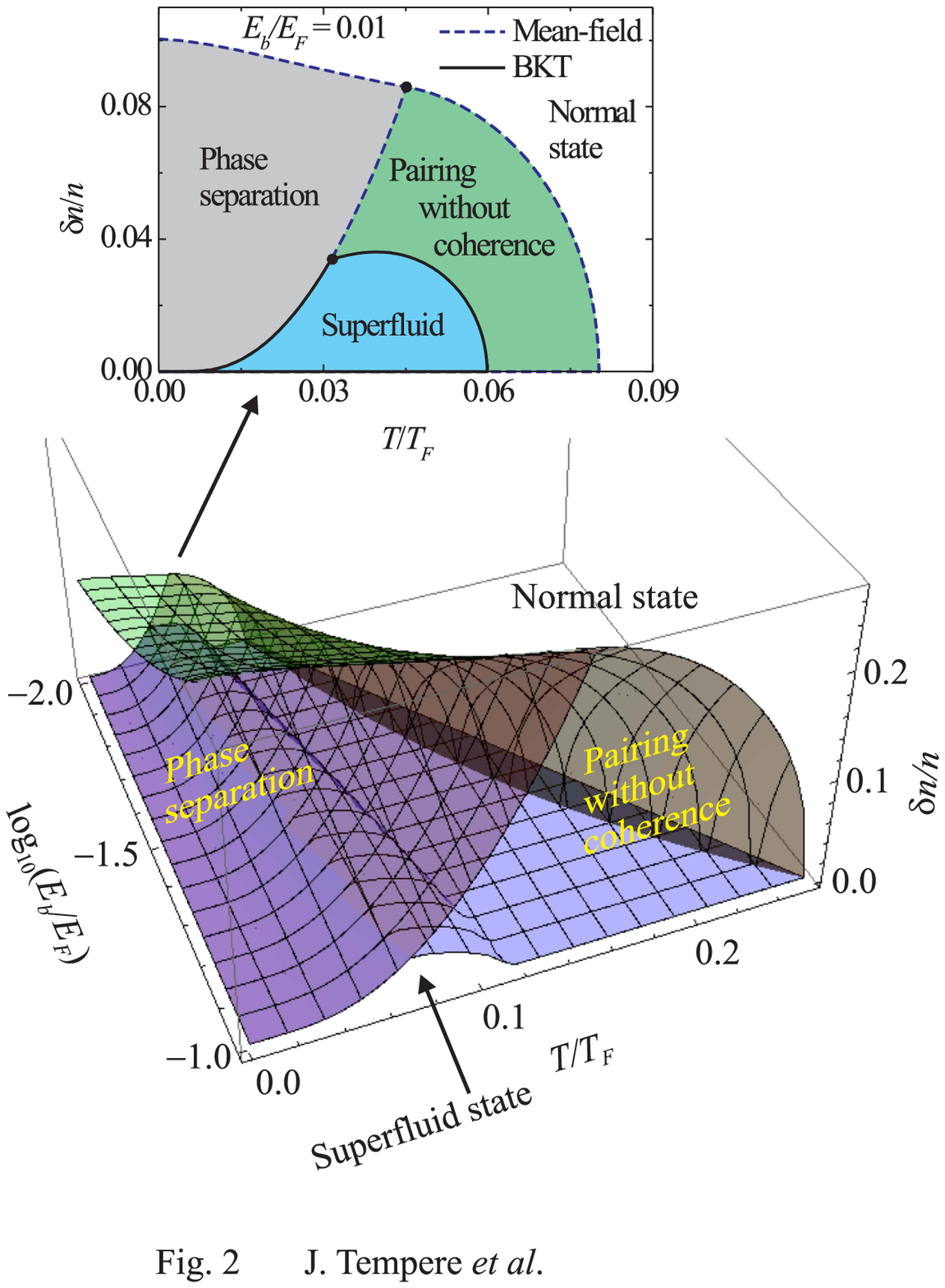}%
\end{center}
\end{figure}

\newpage%

\begin{figure}
[h]
\begin{center}
\includegraphics[
height=4.8992in,
width=3.1047in
]%
{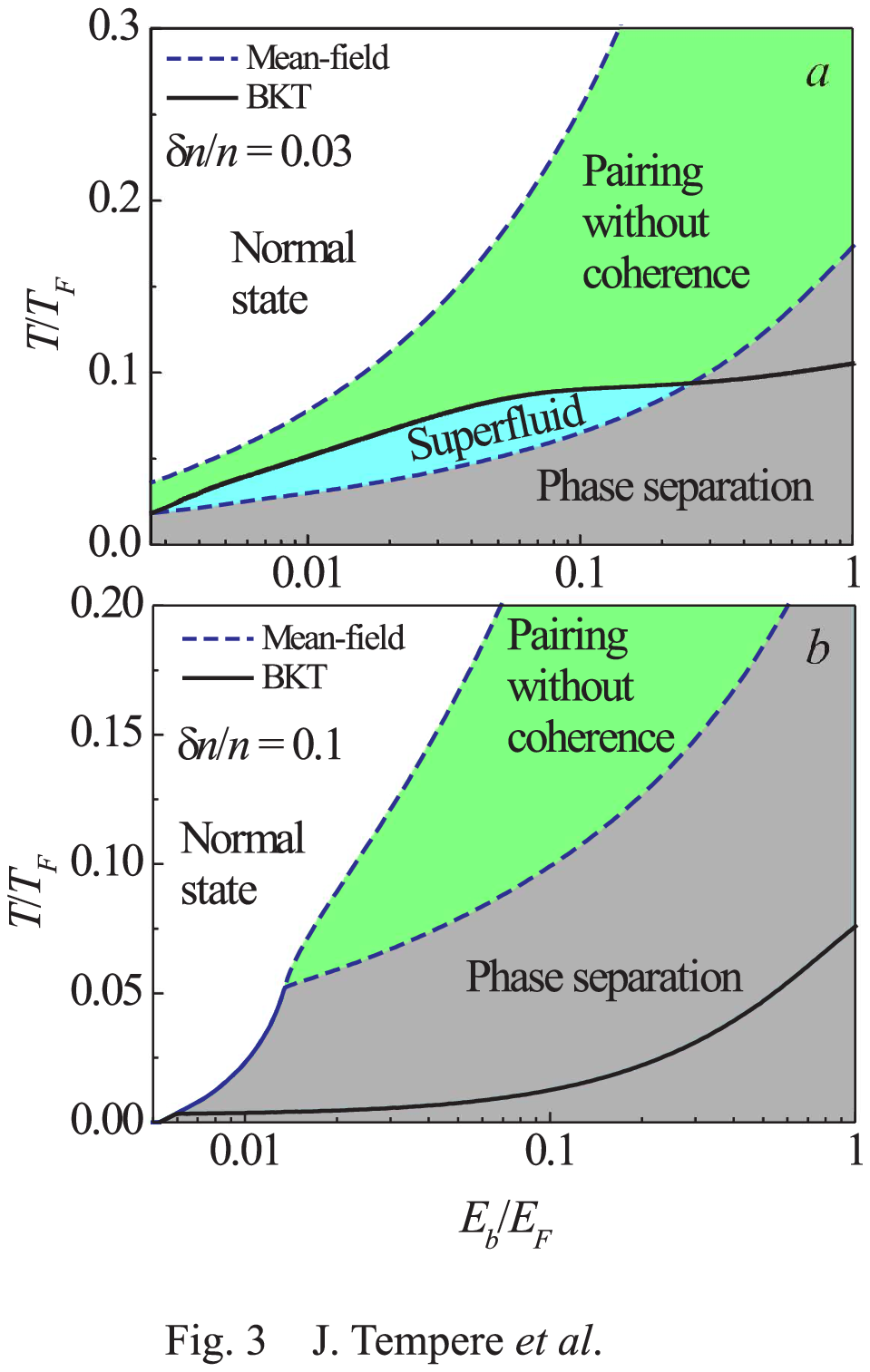}%
\end{center}
\end{figure}

\end{document}